\documentstyle[12pt,epsf,fleqn]{article}

\begin{document}
\begin{titlepage}
\title{
\vspace{-1.5truein}
\begin{flushright}
{\normalsize
{\bf TP-USl/96/17}\\
{\bf December 1996}\\
\vspace{-0.3cm}
{\bf hep-ph/9701358}
}
\end{flushright}
\vspace{0.7truein}
Tau lepton distributions in semileptonic \\
B decays\footnote[1]{ Work supported in part by KBN grants
2P30207607 and PB659/P03/95/08 and by EEC grant ERB-CIPD-CT94-0016.}
}

\author{Marek Je\.{z}abek \\
\normalsize\it
Institute of Nuclear Physics, Cracow, Poland \\
\normalsize\it
Institute of Physics, Silesian University, Katowice, Poland \\
\normalsize\rm and\\
Leszek Motyka\\
\normalsize\it
Institute of Physics, Jagellonian University, Cracow, Poland }
\vspace{2em}
\date{}
\maketitle
\thispagestyle{empty}
\begin{abstract}
Analytic formulae are given for order $\alpha_s$ perturbative QCD
corrections to the double differential distribution of the $\tau$
lepton energy and the invariant mass of the $\tau$~+~antineutrino system
in semileptonic decays of the bottom quark. The corresponding
distribution for B meson decays are obtained by combining these
results and the nonperturbative QCD corrections calculated by Falk
et al. The moments of the energy distribution of the $\tau$ lepton
are calculated.
\end{abstract}
\end{titlepage}
\section{Introduction}
The theoretical description of the semileptonic decays of beautiful hadrons
based on the Heavy Quark Effective Theory (HQET) \cite{VS,PW,IsgW,EH,Grin,Geor}
or the heavy quark mass expansion\cite{CGG,IB,MWB} and on perturbative QCD
seems to be quite satisfactory. The recent ALEPH measurement of the
semileptonic branching ratio for the baryon $\Lambda_b$\cite{MA} indicates that
the problems of the present theory to explain the short lifetime of $\Lambda_b$
are associated with the sector of nonleptonic decays. In contrast to the
nonleptonic widths, the semileptonic widths of beautiful hadrons seem to follow
the theoretical predictions. This conclusion is strengthen by an old observation
that, despite the large differences in the lifetimes, the semileptonic widths of
charmed hadrons are fairly similar. It is clear that the theory of the hadronic
processes involving heavy quarks is still to be improved. Another failure is
the prediction of a large $\Lambda_b$ polarization in $Z^0$ decays. In fact
the measurement\cite{ALEPH2} of $\Lambda_b$ polarization via its semileptonic
decay channels\cite{MeA,CJKK,GR} gives a much smaller value than predicted
in the heavy quark mass limit\cite{CKPS}. All this indicates that further
developments in the theory of nonleptonic processes are expected whereas
the theory of semileptonic decays is firm and stable. Therefore the latter
processes can be used in determination of parameters of the standard model
like the masses $m_b$ and $m_c$ of the charm and bottom quarks and the strong
coupling constant $\alpha_s$ at relatively low scales. Recent
determination\cite{PDG} of the Cabibbo-Kobayashi-Maskawa matrix element
$V_{cb}$ is a spectacular example. It has been argued\cite{CJKK,CJK,MV}
that moments of the lepton spectra can be used in determination of $\alpha_s$,
$m_b$ and $m_c$. Semileptonic decays of the bottom quark are particularly
promising in this respect because in addition to the channels with practically
massless charged leptons (electron and muon) in the final states the channel is
kinematically allowed with the $\tau$ lepton, whose mass is appreciable in
comparison to $m_b$ and $m_c$. This process can be also a window on a new
physics\cite{Kal,GL,GHN}. The theoretical description of inclusive
semileptonic decays of the $b$ quark into the $\tau$ lepton in the framework
of the standard model is therefore necessary. The nonperturbative
QCD corrections to the total rate and the $\tau$ lepton energy spectrum have
been calculated by a number of groups\cite{nonp1,nonp2,nonp3}. Perturbative
QCD corrections to the total semileptonic decay rate are known. In \cite{nonp1}
a numerical approach was used based on an important work by Hokim and
Pham\cite{HP}; see also Bagan et al.\cite{BBBG}. In addition in\cite{CJK}
an analytical formula has been obtained for the distribution of the invariant
mass of the virtual W (i.e. the $\tau\bar{\nu}_\tau$ system) which after performing
one-dimensional numerical integration gives the correction to the semileptonic
rate in perfect agreement with the results of \cite{nonp1}.
The case of perturbative QCD corrections to the $\tau$ lepton energy spectrum
is less satisfactory. In ref.\cite{Boyd} formulae are given for these
corrections as a one-dimensional integral. In the present article we give
another representation of these corrections. We have calculated analytically
the double differential distribution of the $\tau$ energy and the mass of the
$\tau\bar{\nu}_\tau$ system. In the limit $m_\tau\to 0$ our formulae reduce to the
well-established results for the massless case\cite{JK,CJ}. After numerical
integration over the energy of $\tau$ they are in perfect agreement with the
analytical formula given in\cite{CJK}. We were unable, however, to find
agreement between our results and those of ref.\cite{Boyd} for the corrections
to the $\tau$ energy distribution. Our numerical results for the $\tau$
energy distribution have been given in \cite{JM1}. In the present article we
calculate the moments of this distribution.

The article is organized as follows: in Sec.2 kinematical variables are defined
which are used throughout. In Sec.3 the calculation of QCD corrections is
briefly described. In Sec.4 we give the analytic result for the double
differential distribution of the $\tau$ energy and the invariant mass of the
$\tau\bar{\nu}_\tau$ system. This is the main result of the paper. In Sec.5 the
moments of the $\tau$ lepton energy distribution are calculated numerically and
tabulated. In Sec.6 the results of this article are summarized.

\section{Kinematics}
\subsection{Kinematical variables}

The purpose of this section is to define the kinematical variables which are
used in this paper. We describe also the constraints imposed on these variables
for three and four-body decays of the heavy quark.

The calculation is performed in the rest frame of the decaying $b$ quark. Since
the first order perturbative QCD corrections to the inclusive process are taken
into account, the final state can consist either of a produced quark $c$, a
lepton $\tau$ and an anti-neutrino $\bar{\nu}_\tau$ or of the three particles and
a real gluon. The four-momenta of the particles are denoted in the following
way: $Q$ for the $b$~quark, $q$ for the $c$~quark, $\tau$ for the charged
lepton, $\nu$ for the corresponding anti-neutrino and $G$ for the real gluon. By
the assumption that all the particles are on-shell, the squares of their
four-momenta are equal to the squares of masses:
\begin{equation}
Q^2 = m_b ^2, \qquad q^2 = m_c ^2 , \qquad \tau^2 = m_\tau ^2 , \qquad
\nu^2 = G^2 = 0.
\end{equation}
The four-vectors $P = q + G$ and $W = \tau + \nu$ characterize
the quark--gluon system and a virtual intermediating $W$ boson respectively.
We define a set of variables scaled in the units of mass of
the heavy quark $m_b$:
\begin{equation}
\rho = {m_c ^2 \over m_b ^2} , \qquad \eta = {m_\tau ^2 \over m_b ^2},\qquad
x = {2 E_\tau \over m_b ^2}, \qquad t = {W^2 \over m_b ^2}, \qquad
z = {P^2 \over m_b ^2}.
\end{equation}
We choose $m_b$ as the unit of mass (i.e. $Q^2=1$) and
introduce light-cone variables describing the charged lepton:
\begin{equation}
\tau_\pm = {1\over 2} \left( x \pm \sqrt{x^2 - 4\eta} \right).
\end{equation}
The system of the $c$ quark and real gluon is characterized by the following
quantities:
\begin{eqnarray}
P_0 (z)   &=& {1 \over 2}(1-t+z),   \nonumber \\
P_3 (z)   &=& \sqrt{P_0 ^2 - z} = {1\over 2} [1+t^2+z^2-2(t+z+tz)]^{1/2},
                                    \nonumber \\
P_\pm (z) &=& P_0 (z) \pm P_3 (z),  \nonumber \\
{\cal Y}_p (z) &=& {1 \over 2} \ln {P_+ (z) \over P_- (z) } =
                 \ln { P_+ (z) \over \sqrt{z} },
                                    \nonumber \\
\end{eqnarray}
where $P_0 (z)$ and $P_3 (z)$ are the energy and the length of the momentum
vector of the system in the $b$ quark rest frame, ${\cal Y}_p (z)$
is the corresponding rapidity.
Similarly for the virtual boson $W$:
\begin{eqnarray}
W_0 (z) &=& {1 \over 2}(1+t-z),  \nonumber \\
W_3 (z) &=& \sqrt{W_0 ^2 - t} = {1\over 2} [1+t^2+z^2-2(t+z+tz)]^{1/2},  \nonumber \\
W_\pm (z) &=& W_0 (z) \pm W_3 (z),   \nonumber \\
{\cal Y}_w (z) &=& {1 \over 2} \ln {W_+ (z) \over W_- (z) } =
                 \ln { W_+ (z) \over \sqrt{t} },   \nonumber \\
\end{eqnarray}
>From kinematical point of view the three body decay is a special case of the
four body one with the vanishing gluon four-momentum, what is equivalent to a
substitution $z=\rho$ in the previous formulae.
It is convenient to use in this case the following variables:
\[
p_0 = P_0 (\rho) = {1\over 2} (1 - t + \rho ), \qquad
p_3 = P_3 (\rho) = \sqrt{p_0 ^2 - \rho},
\]
\[
p_\pm = P_\pm (\rho) = p_0 \pm p_3,                 \qquad
w_\pm = W_\pm (\rho) = 1 - p_\mp ,
\]
\begin{equation}
Y_p  = {\cal Y}_p (\rho ) = {1 \over 2} \ln {p_+ \over p_-}, \qquad
Y_w  = {\cal Y}_w (\rho ) = {1 \over 2} \ln {w_+ \over w_-}.
\end{equation}
We express also the scalar products which appear in the calculation by the
variables $x$, $t$ and $z$:
\begin{equation}
\begin{array}{ll}
Q \!\cdot\! P \,= {1\over 2}   (1+z-t), & \tau\!\cdot\!\nu \, = {1\over 2}(t - \eta), \\
Q \!\cdot\! \nu \, = {1\over 2} (1-z-x+t),\hspace{3em} &
                                     \tau\!\cdot\! P \, = {1\over 2}(x-t-\eta),   \\
Q \!\cdot\! \tau \, = {1\over 2} x,    & \nu\!\cdot\!\tau \, = {1\over 2}(1-x-z+\eta).\\
\end{array}
\end{equation}
All of the written above products are scaled in the units of the mass of
the $b$~quark.

\subsection{Kinematical boundaries}

The allowed ranges of $x$ and $t$ for the three-body decay
are given by inequalities:
\begin{equation}
2\sqrt{\eta} \leq x \leq 1 + \eta - \rho = x_{\rm \small max}, \qquad
\label{xbound}
\end{equation}
\begin{equation}
t_1 = \tau_- \left( 1 -  {\rho \over 1 - \tau_-} \right)
\leq t \leq
\tau_+ \left( 1 -  {\rho \over 1 - \tau_+} \right) = t_2
\end{equation}
(a region A).
In the case of the four-body process the available region of the phase space
is larger than the region~A. The additional, specific for the four body decay area
of the phase space is denoted as a region~B. Its boundaries are the following:
\begin{equation}
2\sqrt{\eta} \leq x \leq x_{\rm \small max}, \qquad
\eta \leq t \leq t_1.
\label{xyps}
\end{equation}
We remark, that if the charged lepton mass tends to zero than
the region B vanishes.

One can also parameterize the kinematical boundaries of $x$ as
functions of~$t$. In this case we obtain for the region A:
\begin{equation}
\eta  \leq  t  \leq  (1-\sqrt{\rho})^2, \qquad
w_- + {\eta \over w_-} \leq x \leq w_+ + {\eta \over w_+},
\label{yxps}
\end{equation}
and for the region B:
\begin{equation}
\eta  \leq  t  \leq  \sqrt{\eta} \left( 1 - {\rho\over 1 - \sqrt{\eta}}
                                                  \right), \qquad
2\sqrt{\eta}  \leq  x  \leq  w_- + {\eta \over w_-}.
\end{equation}

The upper limit of the mass squared of the $c$-quark --- gluon system is
in the both regions given by
\begin{equation}
z_{\rm\small max} = (1-\tau_+)(1-t/\tau_+),
\end{equation}
whereas the lower limit depends on a region:
\begin{equation}
z_{\rm\small min} = \left\{
\begin{array}{ll}
\rho                  & \mbox{\rm in the region A}, \\
(1-\tau_-)(1-t/\tau_-)   & \mbox{\rm in the region B}. \\
\end{array} \right.
\end{equation}

\section{Calculation of QCD corrections}

The QCD corrected differential rate for
$b \rightarrow c + \tau^- + \bar{\nu}$
reads:
\begin{equation}
d\Gamma = d\Gamma_0 + d\Gamma_{1,3} + d\Gamma_{1,4},
\end{equation}
where
\begin{equation}
d\Gamma_0 = G_F ^2 m_b ^5 |V_{\rm\small CKM} | ^2 {\cal M}_{0,3} ^-
           d{\cal R}_3 (Q;q,\tau,\nu) / \pi^5
\end{equation}
in Born approximation,
\begin{equation}
d\Gamma_{1,3} = {2 \over 3}\alpha_s G_F ^2 m_b ^5 |V_{\rm\small CKM} |^2 {\cal M}_{1,3} ^-
           d{\cal R}_3 (Q;q,\tau,\nu) / \pi^6
\end{equation}
comes from the interference between the virtual gluon contribution
and Born amplitudes, and
\begin{equation}
d\Gamma_{1,4} = {2 \over 3}\alpha_s G_F ^2 m_b ^5 |V_{\rm\small CKM}  |^2 {\cal M}_{1,4} ^-
           d{\cal R}_4 (Q;q,\tau,\nu) / \pi^7
\end{equation}
describes a real gluon emission. $V_{\rm\small CKM} $ is the
Cabibbo--Kobayashi--Maskawa matrix element associated
the $b$ to $c$ or $u$ quark weak transition.
Lorentz invariant $n$-body phase space is defined as
\begin{equation}
d{\cal R}_n(P;p_1, \ldots , p_n ) =
\delta^{(4)} (P - \sum p_i) \prod_i { d^3 {\bf p}_i \over 2 E_i}.
\end{equation}
In Born approximation the rate for the decay into three body final state is
proportional to the expression
\begin{equation}
{\cal M}_{0,3} ^- = F_0 (x,t) =
4 q \!\cdot\! \tau \; Q \!\cdot\! \nu \;= (1 - \rho - x + t )(x - t - \eta),
\end{equation}
where the quantities describing the $W$~boson propagator are neglected.
The three-body phase space is parameterized by the use of Dalitz variables:
\begin{equation}
d{\cal R}_3 (Q;q,\tau,\nu) =
{\pi^2 \over 4} dx \, dt.
\label{Dalitz}
\end{equation}
The matrix element ${\cal M}_{1,3} ^-$ was evaluated in the 
fifties\cite{Behrends} and has the following form:
\begin{eqnarray}
{\cal M}_{1,3} ^- & = &
        - [
     H_0 \; q \!\cdot\!\tau\; Q \!\cdot\! \nu  \;
   + H_+ \; \rho\; Q\!\cdot\!\nu\; Q\!\cdot\!\tau\;
   + H_-\; q \!\cdot\! \nu\; q \!\cdot\! \tau \;
     \nonumber\\
  & &    + {1 \over 2} \rho ( H_+ + H_-) \; \nu \!\cdot\! \tau \;
         + {1 \over 2} \eta \rho  ( H_+ - H_- + H_L ) \; Q \!\cdot\! \nu \;
     \nonumber\\
  & &    - {1 \over 2} H_L \eta \; q \!\cdot\! \nu \; ] ,
\end{eqnarray}
where
\begin{eqnarray}
H_0 & = & 4(1-Y_p p_0/p_3 ) \ln \lambda_G + (2p_0/p_3)
          \left[ {\rm Li}_2 \left( 1 - {p_- w_- \over p_+ w_+ } \right) \right.
    \nonumber\\
    &   & - \left. {\rm Li}_2 \left( 1 - {w_- \over w_+} \right) - Y_p (Y_p+1) +
          2(\ln\sqrt\rho + Y_p)(Y_w + Y_p) \right]
    \nonumber\\
    &   & + [2p_3 Y_p + (1 - \rho - 2t) \ln \sqrt \rho ] / t + 4,
    \nonumber\\
H_\pm & = & {1 \over 2} [ 1 \pm (1-\rho) / t ] Y_p / p_3 \pm
                    {1 \over t} \ln\sqrt\rho,
    \nonumber \\
H_L & = &  {1 \over t} ( 1 -\ln\sqrt\rho) + {1- \rho \over t^2}
           \ln\sqrt\rho + {2 \over t^2} Y_p p_3 + {\rho\over t}
            {Y_p \over p_3}.
\end{eqnarray}
The virtual correction in the interference term is renormalized and 
hence ${\cal M}_{1,3}$ is ultraviolet convergent.
However, the infrared divergences are left. They are regularized by
a small mass of gluon denoted by $\lambda_G$.
According to Kinoshita--Lee--Naunberg theorem, the infrared divergent part
should cancel with the infrared contribution of the four-body decay amplitude
integrated over suitable part of the phase space.

The rate from real gluon emission is proportional to
\begin{equation}
{\cal M}^- _{1,4} =
                    {{\cal B}^- _1 \over (Q\!\cdot\! G)^2 } -
                    {{\cal B}^- _2 \over Q\!\cdot\! G \; P\!\cdot\! G} +
                    {{\cal B}^- _3 \over (P\!\cdot\! G)^2 } ,
\end{equation}
where
\begin{eqnarray}
 {\cal B}_1 ^- & = & \,
  q \!\cdot\! \tau \; [\, Q \!\cdot\! \nu\; (Q \!\cdot\! G \, - 1) + \, G \!\cdot\! \nu \,  -
            \, Q \!\cdot\! \nu\; Q \!\cdot\! G \, +\, G \!\cdot\! \nu\; Q \!\cdot\! G \, ],
  \nonumber\\
 {\cal B}_2 ^- & = & \,
  q \!\cdot\! \tau \; [\, G \!\cdot\! \nu \; Q \!\cdot\! q \, - \, q\!\cdot\!\nu \; Q\!\cdot\! G\, +
  \, Q \!\cdot\! \nu \; (\, q \!\cdot\! G \, - \, Q \!\cdot\! G\, - 2\, q \!\cdot\! Q\,)]
  \nonumber\\
               &   & + \, Q \!\cdot\! \nu \; (\, Q \!\cdot\! \tau \; q \!\cdot\! G \,
                     - \, G \!\cdot\! \tau\;  q \!\cdot\! Q\, ),
  \nonumber\\
 {\cal B}_3 ^- & = &   Q \!\cdot\! \nu \; (\, G \!\cdot\! \tau \; q \!\cdot\! G \, -
                       \rho \; \tau \!\cdot\! P \, ).
  \nonumber\\
\end{eqnarray}
The four-body phase space is decomposed as follows:
\begin{equation}
d{\cal R}_4 (Q;q,G,\tau,\nu) = dz \, d{\cal R}_3 (Q; P,\tau ,\nu)\,
                                     d{\cal R}_2 (P; q,G).
\end{equation}
The four-momentum of the $c$ quark is substituted by $P-G$ and the
integration of ${\cal M}_{1,4} ^-$ over $d{\cal R}_2 (P;q,G)$ is
performed. Lorentz invariance allows to reduce all of appearing
integrals to scalar ones:
\begin{equation}
I_n = \int d{\cal R}_2 (P;q,G)(Q\!\cdot\! G)^n.
\end{equation}
The formulae for $I_n$ have been presented in \cite{CJ}.

In the next step Dalitz parametrization (\ref{Dalitz})
of the three-body phase space $d{\cal R}_3 (Q; P,\tau,\nu)$ is employed.
The expression obtained after integrations has one part
\begin{equation}
{\rm const}\; F_0 (x,t) I_{\rm div},
\end{equation}
where
\begin{equation}
I_{\rm div} = I_{-2} - (1-t+\rho)I_{-1}/(P\!\cdot\! G) +
\rho I_0 / (P \!\cdot\! G)^2,
\end{equation}
which is infrared divergent when $(x,t)$ belongs to the region A
and the rest which is infrared finite.
Since the method used in this calculation is exactly the same as this used in
the previous ones\cite{JK,CJ}, we do not go into the details. The important
point is, that the infrared divergent part is regularized by a small gluon mass
$\lambda_G$ and then contains terms proportional to $\ln \lambda_G$.
The infrared divergent terms from three- and four-body contributions to the
decay rate cancel out and the limit of vanishing gluon mass
$\lambda_G\rightarrow 0$
is performed. This procedure yields well defined double differential
distributions of lepton spectra which are described below.

\section{Analytic results}

The double differential unpolarized distribution of the $\tau$ energy and
$\tau\bar\nu$ invariant mass squared from the $b$ quark decay with first order
perturbative QCD corrections reads \footnote{A Fortran version of the formulae
given in this section is available upon request from 
jezabek@hpjmiady.ifj.edu.pl or leszekm@thp1.if.uj.edu.pl.}
\begin{equation}
{d\Gamma \over dx\, dt} =
\left\{
\begin{array}{ll}
12 \Gamma_0 \left[ F_0 (x,t) - {2\alpha_s \over 3\pi } F_1 ^A (x,t) \right] &
\mbox{for $(x,t)$ in A}, \\
12 \Gamma_0 {2\alpha_s \over 3\pi} F_1 ^B (x,t) & \mbox{for $(x,t)$ in B}, \\
\end{array}
\right.
\label{main}
\end{equation}
where
\begin{equation}
\Gamma_{0} =  {G_F ^2 m_b ^5 \over 192\pi^3} |V_{\rm\small CKM}  |^2,
\end{equation}
\begin{equation}
F_0 (x,t) = (1 - \rho - x + t )(x - t - \eta)
\end{equation}
and
\begin{equation}
F_1 ^A (x,t) = F_0 \Phi_0 + \sum_{n=1}^5 D_n ^A \Phi_n + D_6 ^A,
\label{f1a}
\end{equation}
\begin{equation}
F_1 ^B (x,t) = F_0 \Psi_0 + \sum_{n=1}^5 D_n ^B \Psi_n + D_6 ^B.
\label{f1b}
\end{equation}
The factor of 12 in the formula (\ref{main}) is introduced to meet widely used
\cite{CJK,nonp1,MV} convention for $F_0 (x)$ and $\Gamma_0$.
The symbols present in (\ref{f1a},\ref{f1b}) are defined as follows
\begin{eqnarray}
\Phi_0 &=& {2p_0 \over p_3}
           \left[
           {\rm Li}_2~\left( 1-{1-\tau^+    \over p_+} \right)
         + {\rm Li}_2~\left( 1-{1-t/\tau^+  \over p_+} \right)
           \right.
   \nonumber\\
     &&  \quad  - {\rm Li}_2~\left(1-{1-\tau^+   \over p_-} \right)
          - {\rm Li}_2~\left(1-{1-t/\tau^+ \over p_-} \right)
   \nonumber\\
     &&  \quad \left.
          + {\rm Li}_2~(w_-) - {\rm Li}_2~(w_+) + 4Y_p~\ln\sqrt{\rho}
          \right]
   \nonumber\\
     &&  \quad
         + 4\left(1-{p_0\over p_3} Y_p \right) \ln~(z_{\rm\small max}-\rho)-
           4\ln z_{\rm\small max},
   \nonumber\\
\Phi_1 &=&
         {\rm Li}_2~(w_-) + {\rm Li}_2~(w_+) - {\rm Li}_2~(\tau_+) - {\rm Li}_2~(t/\tau_+),
         \nonumber\\
\Phi_2 &=& {Y_p \over p_3},                \nonumber\\
\Phi_3 &=& {1\over 2} \ln\sqrt{\rho},   \nonumber\\
\Phi_4 &=& {1\over 2} \ln (1-\tau_+),      \nonumber\\
\Phi_5 &=& {1\over 2} \ln~(1-t / \tau_+ ),
\end{eqnarray}

\begin{eqnarray}
\Psi_0  &=&  4\left( {p_0\over p_3} Y_p - 1 \right)
       \ln~\left(
       { z_{\rm\small max} - \rho \over z_{\rm\small min} - \rho}
            \right)
       + 4\ln\left(
       { z_{\rm\small max} \over z_{\rm\small min}}
                 \right)
      \nonumber\\
        & &  +  {2p_0\over p_3}
           \left[
             {\rm Li}_2\left( 1 - {1-\tau_+ \over p_-}     \right)
           + {\rm Li}_2\left( 1 - {1-t/\tau_+ \over p_-}   \right)
           \right.
      \nonumber\\
        & & \qquad
            - {\rm Li}_2\left( 1 - {1-\tau_+   \over p_+}   \right)
            - {\rm Li}_2\left( 1 - {1-t/\tau_+ \over p_+}   \right)
            \nonumber\\
        & & \qquad
            + {\rm Li}_2\left( 1 - {1-\tau_-   \over p_+}   \right)
            + {\rm Li}_2\left( 1 - {1-t/\tau_- \over p_+}   \right)
            \nonumber\\
        & & \qquad \left.
            - {\rm Li}_2\left( 1 - {1-\tau_- \over p_-}     \right)
            - {\rm Li}_2\left( 1 - {1-t/\tau_- \over p_-}   \right)
            \right],
      \nonumber\\
\Psi_1 &=& {\rm Li}_2~(\tau_+) + {\rm Li}_2~(t/\tau_+)
         - {\rm Li}_2~(\tau_-) - {\rm Li}_2~(t/\tau_-),
      \nonumber\\
\Psi_2 &=&  {1\over 2}\ln (1-\tau_-),            \nonumber\\
\Psi_3 &=&  {1\over 2}\ln (1- t / \tau_-),       \nonumber\\
\Psi_4 &=&  {1\over 2}\ln (1 - \tau_+),          \nonumber\\
\Psi_5 &=& {1\over 2}\ln (1 - t / \tau_+).
\end{eqnarray}
We introduce $C_1 \ldots C_5$ to simplify the formulae for $D_n ^A$ and
$D_n ^B$:
\begin{eqnarray}
C_1 &=& - 2xt + x + t + t^2 - \eta x + \eta t + 5\eta+\rho (x-t-\eta),
    \nonumber\\
C_2 &=& \sqrt{x^2 - 4\eta}\; (6 - 2x + 2t - t^2/\eta  + \eta  - 2\rho ),
    \nonumber\\
C_3 &=& - 5 - x + 2x^2 - 3t + 5t^2 - 6xt +xt^2/\eta- 2\eta x+\eta t
    \nonumber\\
    & & +11\eta - \eta ^2 + \rho  ( 4 + 7x - 11t - 7\eta  ) + \rho^2,
    \nonumber\\
C_4 &=& - 4xt - 3x + 2x^2 + 11t + 2t^2 - xt^2/\eta + 2x\eta /t- 2\eta x
    \nonumber\\
    & & -10\eta /t + 3\eta t + 3\eta + 5\eta ^2/t^2 - 6\eta ^2/t +
           \rho ( 5x - 7t
    \nonumber\\
    & & + 2x\eta /t+ 8\eta /t - 11\eta - 4\eta^2/t^2) +
    \rho^2 ( 2 - \eta /t )\eta /t,
    \nonumber\\
C_5 & = & {1 \over 2}\sqrt{x^2 - 4\eta}\;
    \left[
     5 - 2t^2/\eta  - 5\eta /t - \eta  + 3t - 3\rho (1 - \eta /t)
    \right.
    \nonumber\\
    & & \left. + {\eta \rho \over t}
        { (1+\rho)(1 - \eta /t) \over x-t-\eta /t}
        + {\rho ( t - \eta  ) \over 1-x+\eta }
        \right],
\label{Cn}\\
%\end{eqnarray}
%The expressions $D_n ^A$ and $D_n ^B$ read
%\begin{eqnarray}
   D_1 ^A & = &  C_1,
          \nonumber\\
   D_2 ^A & = &
         ( - 5 +2x - 3t + \rho ) p_3^2 - 2\rho t
           + 2\eta \{
              - 4p_3^4/t^2 + [ x - 7t
          \nonumber\\
       & & + t^2 - \rho(  x + t ) ] p_3^2/t^2
                    \}
           + \eta ^2 [( 3 + t + \rho ) p_3^2/t^2 + 2\rho/t ] ,
          \nonumber\\
   D_3 ^A & = &
           - 4p_3^2 - 4 - 2xt + 2x + 4t^2 + 2\rho( 2 + 3x - 7t )
          \nonumber\\
       & & + 2\eta [ 4 ( x - 1 + \rho )p_3^2/t^2
           + 6 + x/t - x - 6/t
          \nonumber\\
       & & + \rho  (  - 6 + x/t + 6/t ) ]
           + 2\eta ^2 [ - 2p_3^2 + 2 - 2t - \rho  ( 2 + t ) ] /t^2,
          \nonumber\\
   D_4 ^A & = & -C_2 - C_3,  \nonumber\\
   D_5 ^A & = &  C_2 - C_4,  \nonumber\\
   D_6 ^A & = &
        {1 \over 4}( 5x + 9xt  + 4t - 6t^2  - 2xt^2/\eta - \eta x/t +
                                          \eta x - 4\eta /t - 2\eta t
           \nonumber\\
       &   & - 22\eta  + 6\eta ^2/t ) + {1 \over 4}\rho
                ( - 3x + 5t - x\eta /t + 9\eta /t + 5\eta
           \nonumber\\
       &   &  - \eta ^2/t^2 + 2\eta ^2/t ) - {1 \over 4}
            { \rho (t - \eta ) (1 - \eta ) \over 1-x+\eta}
            - {1 \over 4} {\rho^2 \eta \over t}  ( 3 + \eta /t )
           \nonumber\\
       &   &    - {1 \over 4} {\rho \eta \over t}
                 {( 1 + \rho )( t - \eta /t )( 1 - \eta /t )
                  \over x-t-\eta /t}    - {1 \over 2}  C_5,
\label{Dn}
\end{eqnarray}

\begin{eqnarray}
D_1 ^B & = & C_1,        \nonumber\\
D_2 ^B & = & C_2 - C_3,  \nonumber\\
D_3 ^B & = & -C_2 - C_4, \nonumber\\
D_4 ^B & = & C_2 + C_3,  \nonumber\\
D_5 ^B & = & -C_2 + C_4, \nonumber\\
D_6 ^B & = &  C_5.
\label{dprzezc}
\end{eqnarray}

One can perform the limit $\rho \rightarrow 0$, what corresponds to the
decay of the bottom quark to an up quark and leptons.
The formulae, which are much simpler in this case,
are presented in the same manner as previously the full results:

\begin{equation}
{d\widetilde{\Gamma} \over dx\, dt} =
\left\{
\begin{array}{ll}
12 \Gamma_0 \left[ \widetilde{F}_0 (x,t) -
                 {2\alpha_s \over 3\pi } \widetilde{F}_1 ^A (x,t) \right] &
\mbox{for $(x,t)$ in A}, \\
12 \Gamma_0 {2\alpha_s \over 3\pi} \widetilde{F}_1 ^B (x,t)
                 & \mbox{for $(x,t)$ in B}, \\
\end{array}
\right.
\label{mainmlss}
\end{equation}
where
\begin{equation}
\widetilde{F}_0 (x,t) = (1 - x + t )(x - t - \eta)
\end{equation}
and
\begin{equation}
\widetilde{F}_1 ^A (x,t) =
\widetilde{F}_0 \widetilde{\Phi}_0 +
        {\cal D}_1 ^A \widetilde{\Phi}_1
      + {\cal D}_{2\diamond 3} ^A \widetilde{\Phi}_{2 \diamond 3}
      + {\cal D}_4 ^A \widetilde{\Phi}_4
      + {\cal D}_5 ^A \widetilde{\Phi}_5
      + {\cal D}_6 ^A, \nonumber
\end{equation}
\begin{equation}
\widetilde{F}_1 ^B (x,t) =
\widetilde{F}_0  \widetilde{\Psi}_0 + \sum_{n=1}^5
        {\cal D}_n ^B \widetilde{\Psi}_n + {\cal D}_6 ^B,
\end{equation}
where
\begin{eqnarray}
\widetilde{\Phi}_0 & = &
                     2 \left[ {\rm Li}_2 \left( {\tau_+ - t \over 1 - t} \right) +
                     {\rm Li}_2 \left(  {1/\tau_+ - 1 \over 1/t - 1 } \right)
                   + {\rm Li}_2 (t) \right] + {1\over 2}\pi^2
\nonumber\\
           &&      + \ln^2 (1-\tau_+) + 2 \ln^2 (1-t) + \ln^2 ( 1- t/\tau_+ )
\nonumber\\
           &&        -2 \ln (1-t) \ln z_{\rm \small max},
\nonumber\\
\widetilde{\Phi}_1  & = & {\pi ^2 \over 12} + {\rm Li}_2 (t) - {\rm Li}_2 (\tau_+) -
                         {\rm Li}_2 (t/\tau_+),
            \nonumber\\
\widetilde{\Phi}_{2\diamond 3} & = & {2 \ln (1-t) \over 1 - t},
            \nonumber\\
\widetilde{\Phi}_4 & = & \Phi_4, \nonumber\\
\widetilde{\Phi}_5 & = & \Phi_5
\end{eqnarray}
and
\begin{eqnarray}
\widetilde{\Psi}_0 & = & 2 \left[
                 {\rm Li}_2 \left(  {\tau_- - t \over 1 - t}  \right)
               + {\rm Li}_2 \left( {1/\tau_- - 1 \over 1/t-1} \right)
               - {\rm Li}_2 \left( \tau_+ - t \over 1 - t     \right)
                 \right.
\nonumber\\
     &&          \left.
               - {\rm Li}_2 \left( 1/\tau_+ - 1 \over 1/t - 1 \right)
               + \ln (1-t) \ln\left(
                   {z_{\rm \small max} \over z_{\rm\small min}}
                \right)
          \right]
\nonumber\\
     &&       - \ln \left({1-\tau_+ \over 1 - \tau_-} \right)
                \ln [(1-\tau_+)(1-\tau_-)]
\nonumber\\
     &&       - \ln \left( {1 - t/\tau_+ \over 1 - t/\tau_- } \right)
                \ln [ (1 - t/\tau_+)(1 - t/\tau_-) ],
\nonumber\\
\widetilde{\Psi}_n & = & \Psi_n \qquad n=1\ldots 5 .
\end{eqnarray}
The formulae for ${\cal C}_n$ and ${\cal D}_n ^A$
are more compact than (\ref{Cn}) and (\ref{Dn}):
\begin{eqnarray}
     {\cal C}_1 & = & - \eta x + \eta t + 5\eta - 2xt + x + t + t^2,
\nonumber\\
     {\cal C}_2 & = & \sqrt{x^2-4\eta}\;( 6 - t^2/\eta + \eta - 2x + 2t ),
\nonumber\\
     {\cal C}_3 & = & - 5 + xt^2/\eta - 2\eta x + \eta t + 11\eta - \eta^2
               - 6xt - x + 2x^2 - 3t + 5t^2,
\nonumber\\
     {\cal C}_4 & = & - xt^2/\eta + 2\eta x / t - 2\eta x - 10 \eta / t
                     + 3\eta t + 3\eta + 5\eta^2/t^2 - 6\eta^2/t
\nonumber\\
                    &   & - 4xt - 3x + 2x^2 + 11t + 2t^2,
\nonumber\\
     {\cal C}_5 & = & {1\over 2} \sqrt{x^2 - 4\eta}
                \left( 5 - 2 t^2/\eta - 5\eta /t - \eta + 3 t \right),\\
%\end{eqnarray}
%
%\begin{eqnarray}
   {\cal D}_1 ^A & = & {\cal C}_1 ,
\nonumber\\
   {\cal D}_{2 \diamond 3}^A & = &
       - {5\over 4} + {1\over 2}\eta x/t^2 - \eta x / t + {1\over 2}\eta x -
         {1\over 2}\eta /t^2 - {3\over 2} \eta / t - {5\over 2}\eta t +
         {9\over 2} \eta
\nonumber\\
             & & + {3\over 4}\eta^2 / t^2  - {5\over 4}\eta^2/t +
                   {1\over 4}\eta^2 t + {1\over 4}\eta^2
                    - xt + {1\over 2} xt^2 + {1\over 2} x + {7\over 4} t
\nonumber\\
             & &  + {1\over 4}t^2 - {3\over 4}t^3 ,
\nonumber\\
   {\cal D}_4 ^A & = & \sqrt{x^2 - 4\eta}\;
           (- 6 + t^2/\eta - \eta + 2x - 2t ) + 5 - xt^2/\eta  +2\eta x
\nonumber\\
           & &     - \eta t - 11\eta +  \eta^2 + 6xt + x - 2x^2 + 3t - 5t^2 ,
\nonumber\\
   {\cal D}_5 ^A & = &\sqrt{x^2-4\eta}\; ( 6 - t^2/\eta + \eta - 2x + 2t )
            + xt^2/\eta - 2\eta x/t + 2\eta x
\nonumber\\
         &  & + 10\eta / t - 3\eta t - 3\eta - 5\eta^2/ t^2 + 6\eta^2 / t
              + 4xt + 3x - 2x^2
\nonumber\\
         &  & - 11t - 2t^2 ,
\nonumber\\
   {\cal D}_6 ^A & = & \sqrt{x^2-4\eta}\; \left(
     - {5\over 4} + {1\over 2} t^2/\eta +
       {5\over 4}\eta / t + {1\over 4}\eta - {3\over 4} t \right) -
       {1\over 2} xt^2 /\eta 
\nonumber\\
      & &  - {1\over 4}\eta x/t
           + {1\over 4}\eta x - \eta/t - {1\over 2}\eta t - {11\over 2} \eta
           + {3\over 2}\eta^2 /t + {9\over 4} xt + {5\over 4} x
\nonumber\\
      & &    + t - {3\over 2} t^2  .
\end{eqnarray}
The coefficients ${\cal D}_n ^B$ are expressed by
${\cal C}_n$ in the same way as $D_n ^B$ are expressed
by $C_n$ in the set of equations (\ref{dprzezc}).

The listed above results have been tested by comparison with the calculation made
earlier in simpler cases. One of the cross checks was arranged by fixing the
mass of the produced lepton to zero.
Our results are in this limit algebraically identical
with those for the massless charged lepton\cite{JK,CJ}.
On the other hand one can numerically integrate the calculated
double differential distribution over $x$, with the limits given by the
kinematical boundaries:
\begin{equation}
\int_{2\sqrt{\eta}} ^{w_+ + \eta / w_+ } {d\Gamma\over dx\, dt}
(x,t;\rho,\eta) = {d\Gamma \over dt} (t;\rho,\eta) .
\label{testint}
\end{equation}
Obtained in such a way differential
distribution of $t$ agrees with recently published\cite{CJK}
analytical formula describing this distribution. This test is particularly
stringent because one requires two functions of three variables
($t,\rho$ and $\eta$) to be numerically equal for any values of the
arguments. We remark, that for higher values of $t$
only the region A contributes to the integral (\ref{testint}) and for lower
values of $t$ both the regions~A and B contribute.
This feature of the test is very helpful ---
the formulae for $F_1 (x,t)$, which are different for the
regions A and B can be checked separately. For completeness we quote the
formulae derived in \cite{CJK}:
\begin{eqnarray}
{{\rm d}\Gamma\over {\rm d}t} &=&
\Gamma_{0}\,
\left( 1-{\eta \over t}\right)^2 \,
\left\{
\left( 1+{\eta \over 2t}\right)
\left[ {\cal F}_0(t) - {2\alpha_s\over 3 \pi} {\cal F}_1(t)\right]
\right.
\nonumber\\
 & & \left.
\vspace{20em}
+ {3\eta \over 2t}
\left[ {\cal F}_0^s(t) - {2\alpha_s\over 3 \pi} {\cal F}_1^s(t)\right]
\right\},
\label{maint}
\end{eqnarray}
where
\begin{eqnarray}
{\cal F}_0(t) &=& 4 p_3\,
\left[ \, (1-\rho)^2 + t(1+\rho) - 2 t^2\, \right],
\\
{\cal F}_0^s(t)
&=& 4p_3\,\left[\, (1-\rho)^2 -t ( 1+\rho)\, \right],
\\
{\cal F}_1(t)&=& {\cal A}_1 \Psi  + {\cal A}_2 Y_w
+ {\cal A}_3 Y_p + {\cal A}_4 p_3 \ln\rho + {\cal A}_5 p_3,
\\
{\cal F}_1^s(t)&=& {\cal B}_1 \Psi  + {\cal B}_2 Y_w
+ {\cal B}_3 Y_p + {\cal B}_4 p_3 \ln\rho + {\cal B}_5 p_3
\label{massive}
\end{eqnarray}
and
\begin{eqnarray}
\Psi &=& 8 \ln (2 p_3) -2\ln t\, + \,
\left[ 2{\rm Li}_2 (w_-) -2{\rm Li}_2 (w_+)
 +4 {\rm Li}_2 ({2p_3/ p_+}) \right.
\nonumber \\ && \left.
  -4Y_p \ln({2p_3/ p_+})
-\ln p_- \ln w_+ + \ln p_+ \ln w_- \right]\, 2p_0/p_3,
\end{eqnarray}
\begin{eqnarray}
{\cal A}_1 &=&  {\cal F}_0(t),
\nonumber\\
{\cal A}_2 &=&
- 8 (1-\rho) \left[ 1 +t -4 t^2-
\rho (2-t) +\rho^2 \right],
\nonumber\\
{\cal A}_3 &=&
- 2 \left[ 3 + 6 t   -21 t^2  + 12 t^3
-\rho (1+12t+5t^2)  \right.
\nonumber\\
& & \left. +\rho^2(11+2t) - \rho^3 \right],
\nonumber\\
{\cal A}_4 &=&
- 6 \left[ 1  + 3 t  - 4 t^2
-\rho (4-t) + 3 \rho^2 \right],
\nonumber\\
{\cal A}_5 &=&
- 2 \left[ 5+9t-6t^2 -\rho( 22 - 9t) + 5 \rho^2 \right],
\\
{\cal B}_1 &=&  {\cal F}^s_0(t),
\nonumber\\
{\cal B}_2 &=& -8 (1-\rho)
\left[(1-\rho)^2-t (1+\rho)\right],
\nonumber\\
{\cal B}_3 &=& - 4(1 -  \rho)^4/t
         -2  (-1 + 3  \rho + 15 \rho^2  - 5 \rho^3)
\nonumber\\
           & & + 8  (1 + \rho) t - 6  (1 + \rho) t^2,
\nonumber\\
{\cal B}_4 &=&      -4 (1-\rho)^3 /t
           -2 (1-\rho) (1-11 \rho)
           +6 (1 + 3 \rho) t,
 \nonumber\\
{\cal B}_5 &=&  -6 (1-3 \rho) (3-\rho)
+18 t (1+\rho).
\end{eqnarray}
The formula (\ref{maint}) consists of two parts with
transparent physical meaning\cite{CJK}.
The first term in the curly bracket containing the function ${\cal F}_{0,1}(t)$
corresponds to the real $t \rightarrow bW$ decay in the limit of vanishing
$b$~quark mass. The other one is related to a fictitious decay
$t\rightarrow bH^+$ of a heavy quark $t$
to the quark $b$ and a charged Higgs particle $H^+$. The formula \cite{JK1}
describing the first order QCD correction ${\cal F}_1 (t)$ has been confirmed
by many groups, cf. note added in ref.~\cite{CJ}. On the other hand the formula
for ${\cal F}_1 ^s (t)$ has been obtained only in \cite{CD} and it still
requires an independent check.
Our calculation confirms the result~(51) of ref.~\cite{CD}.

\section{Moments of $\tau$ energy distribution}
It has been argued in \cite{CJKK,CJK,MV} that moments of charged lepton energy
distribution from $b$-quark decay are valuable source of knowledge about the
physical parameters involved in the process and are only weakly affected by 
soft processes. Similar to the moments are the 
observables which are used in \cite{GKLW} to find the values of some 
nonperturbative parameters of HQET.   
 
Thus, in the notation of \cite{MV}
\begin{equation}
M_n = \int_{E_{\rm \small min}} ^{E_{\rm \small max}}
E_\tau ^n {d\Gamma \over dE_\tau} dE_\tau,
\end{equation}
\begin{equation}
r_n = {M_n \over M_0},
\end{equation}
where $E_{\rm \small min}$ and $E_{\rm \small max}$ are the lower and upper
limits for $\tau $ energy and $M_n$ involve both perturbative and
nonperturbative
QCD corrections to $\tau$ energy spectrum. The nonperturbative  $1/m_b ^2$
corrections to the charged lepton spectrum from semileptonic $B$ decays have been
derived in the framework of HQET\cite{nonp1,nonp2,nonp3} and even $1/m_b ^3$
corrections are known\cite{GK96}.
Up to order of $1/m_b^2$ 
the corrected heavy lepton energy spectrum can be written in the following way:
\begin{equation}
{1 \over \Gamma_0} {d\Gamma \over dx} =
f_0(x) -
{2\alpha_s \over 3\pi} f_1 (x) +
{\lambda_1 \over m_b^2} f_{np} ^{(1)} (x) +
{\lambda_2 \over m_b^2} f_{np} ^{(2)} (x),
\end{equation}
where $\lambda_1$ and
$\lambda_2$ are the HQET parameters corresponding to the $b$~quark kinetic
energy and the energy of interaction of the $b$ quark magnetic moment with
chromomagnetic field produced by a light quark in the meson $B$.
The functions $f_{np} ^{(1,2)}$ can be easily extracted from the formula (2.11)
in~\cite{nonp1}.

Following ref.\cite{MV} we expand the ratio $r_n$:
\begin{equation}
r_n = r_n ^{(0)}\left( 1 - {2\alpha_s \over 3\pi}   \delta_n ^{(p)}
                         + {\lambda_1 \over m_b ^2} \delta_n ^{(1)}
                         + {\lambda_2 \over m_b ^2} \delta_n ^{(2)}
                                                                \right),
\label{rn}
\end{equation}
where $r_n ^{(0)}$ is the lowest approximation of $r_n$,
\begin{equation}
r_n ^{(0)} = \left( {m_b \over 2} \right) ^n
{\int _{2\sqrt{\eta}} ^{1+\eta-\rho} f_0 (x) x^n \, dx   \over
 \int _{2\sqrt{\eta}} ^{1+\eta-\rho} f_0 (x) \, dx }.
\end{equation}
Each of the $\delta^{(i)} _n$ is expressed by integrals of the corresponding
correction function $f^{(i)}(x)$ and the tree level term $f_0 (x)$
\begin{equation}
\delta_n ^{(i)} =
{\int _{2\sqrt{\eta}} ^{1+\eta-\rho} f ^{(i)} (x) x^n \, dx   \over
 \int _{2\sqrt{\eta}} ^{1+\eta-\rho} f_0 (x)   x^n \, dx } -
{\int _{2\sqrt{\eta}} ^{1+\eta-\rho} f ^{(i)}   (x) \, dx   \over
 \int _{2\sqrt{\eta}} ^{1+\eta-\rho} f_0 (x)       \, dx },
\label{deltadef}
\end{equation}
where the index $i$ may denote any type of the described above corrections.
The coefficients $\delta_n ^{(i)}$ depend only on two ratios of masses of the
charged lepton and the $c$~quark to the mass $m_b$.
It is convenient to introduce the functional dependence of
the $\delta_n ^{(i)}$ on the masses in the following way:
\begin{equation}
\delta_n ^{(i)} (m_b,m_c,m_\tau) =
\delta_n ^{(i)} \left( {m_b \over m_\tau}, {m_c \over m_b} \right).
\label{deltapar}
\end{equation}
Since the values of $b$ and $c$ quark masses are not known precisely, we
calculated the $\delta$'s in a reasonable range of them. We have chosen
$4.4 \;{\rm GeV} \leq m_b \leq 5.2\; {\rm GeV}$ and
$0.25 \leq m_c / m_b \leq 0.35$. In this range the dependence of the function
$\delta_n ^{(i)} \left( {m_b \over m_\tau}, {m_c \over m_b} \right)$ on the
variables can be approximated by a second order polynomial with the maximal
relative error smaller then 1\% . We propose:
\begin{eqnarray}
\delta (p,q)&=&a + b(p-p_0) + c(q-q_0) + d(p-p_0)^2
\nonumber\\
            & &+e(p-p_0)(q-q_0)+f(q-q_0)^2,
\label{poly}
\end{eqnarray}
where $p=m_b / m_\tau$, $p_0 = 4.75 {\rm GeV} / 1.777 {\rm GeV} = 2.6730$,
$q=m_c / m_b$, $q_0 = 0.28$ and the polynomial coefficients are fitted for each
of the $\delta_n ^{(i)}$ separately. Such a parameterization marks out the
realistic masses of quarks: $m_b = 4.75\; {\rm GeV}$ and
$m_c = 1.35\; {\rm GeV}$, for which the coefficient
$\delta_n ^{(i)} = a_n ^{(i)}$.

The results of the fits of formula (\ref{poly}) to the calculated numerically
coefficients $\delta_n ^{(i)} ( m_b / m_\tau, m_c / m_b)$ are listed in
Table~1 for all the three types of $\delta_n ^{(i)}$ and $n=1\ldots 5$.
With use of these results it is possible to estimate the order of relative
correction to the ratio $r_n$ (\ref{rn}), including reasonable values of
$\alpha_s$, $\lambda_1$ and $\lambda_2$. 
The value of $\lambda_2$ may be easily determined from a measured 
$B$~--~$B^*$ splitting, and $\alpha_s$ should be of the order of 
$\alpha_s(m_b)$.
We assume $\alpha_s \simeq 0.3$,
$\lambda_2 = 0.12 \; {\rm GeV}^2$, keep the mass of the $b$ quark fixed to
4.75~GeV and the mass of the $c$~quark equal to 3.35~GeV.
It was claimed in \cite{GKLW} that $\lambda_1$  may be constrained 
$\lambda_1 = -0.35\pm 0.05$~GeV$^2$ but we choose more conservative estimate
$-0.60 \; {\rm GeV}^2 \leq \lambda_1 \leq -0.15 \; {\rm GeV}^2$.
Thus the perturbative correction to $r_n / r_n ^{(0)}$ is
about~$-0.0015$~($-0.0075$) for $n=1$~($n=5$) whereas the nonperturbative
corrections are larger: for the "kinetic energy" part
$0.008 \pm 0.004$~($0.06 \pm 0.03)$, $n=1$~($n=5$) and for the
"chromomagnetic" one $-0.01$~($-0.05$), $n=1$~($n=5$). Although the two
nonperturbative terms partly cancel each other, 
the perturbative correction to the moments are smaller than the 
nonperturbative contributions.
The ratios $r_n$ may be used to fix the value of $\lambda_1$ if suitable
precise measurements are performed.

\begin{figure}[hbpt]
Table 1. Polynomial coefficients defined by the formula (\ref{poly})
characterizing how the three different types of corrections $\delta_n ^{(i)}$
depend on quark masses. \vspace{1ex} \\

\noindent A. The perturbative correction. \\

\begin {tabular}{|r|r|r|r|r|r|r|}
\hline
                  &   $a$   &   $b$   &   $c$   &   $d$   &   $e$   &  $f$    \\
\hline
$\delta_1 ^{(p)}$ & 0.0213  & 0.0097  & -0.1059 & -0.0061 & -0.0202 & 0.178   \\
\hline
$\delta_2 ^{(p)}$ & 0.0443  & 0.0216  & -0.2213 & -0.0108 & -0.0454 & 0.363   \\
\hline
$\delta_3 ^{(p)}$ & 0.0687  & 0.0355  & -0.3457 & -0.0157 & -0.0808 & 0.554   \\
\hline
$\delta_4 ^{(p)}$ & 0.0946  & 0.0512  & -0.4781 & -0.0193 & -0.1163 & 0.746   \\
\hline
$\delta_5 ^{(p)}$ & 0.1219  & 0.0684  & -0.6168 & -0.0259 & -0.1538 & 0.934   \\
\hline
\end{tabular}                        \\

\vspace{1em}

\noindent
B. The nonperturbative correction corresponding to kinetic energy of the $b$ quark.
\\

\begin {tabular}{|r|r|r|r|r|r|r|}
\hline
                  &   $a$   &   $b$   &  $c$   &   $d$   &   $e$   &   $f$   \\
\hline
$\delta_1 ^{(1)}$ &  -0.500 &   0.0   &  0.0   &   0.0   &   0.0   &   0.0   \\
\hline
$\delta_2 ^{(1)}$ &  -1.105 &  -0.081 & 0.220  &  0.011  &   0.016 &   0.26  \\
\hline
$\delta_3 ^{(1)}$ &  -1.823 &  -0.254 & 0.680  &  0.035  &   0.039 &   0.84  \\
\hline
$\delta_4 ^{(1)}$ &  -2.664 &  -0.526 & 1.396  &  0.074  &   0.064 &   1.77  \\
\hline
$\delta_5 ^{(1)}$ &  -3.633 &  -0.904 & 2.377  &  0.131  &   0.068 &   3.11  \\
\hline
\end{tabular}  \\
%\end{figure}
%\begin{figure}[hbpt]

\vspace{1em}

C. The nonperturbative correction corresponding to the interaction of the $b$~quark
spin with chromomagnetic field.  \\

\begin {tabular}{|r|r|r|r|r|r|r|}
\hline
                  &   $a$   &   $b$   &  $c$   &   $d$   &   $e$   &   $f$   \\
\hline
$\delta_1 ^{(2)}$ &  -1.731 & -0.732  & 2.784  & 0.212   &  -0.498 &  -1.71  \\
\hline
$\delta_2 ^{(2)}$ &  -3.565 & -1.561  & 5.824  & 0.437   &  -0.975 &  -3.45  \\
\hline
$\delta_3 ^{(2)}$ &  -5.496 & -2.473  & 9.069  & 0.681   &  -1.506 &  -5.13  \\
\hline
$\delta_4 ^{(2)}$ &  -7.517 & -3.453  & 12.472 & 0.961   &  -2.141 &  -6.72  \\
\hline
$\delta_5 ^{(2)}$ &  -9.621 & -4.485  & 15.987 & 1.279   &  -2.955 &  -8.21  \\
\hline
\end{tabular}  \\
\end{figure}

\section{Summary}

The results of this article can be summarized in the following way:

\noindent
(i) an analytical expression has been presented for the double differential 
distribution $d\Gamma / dx\, dt$ of the $\tau$~lepton energy and the
invariant mass of the leptons in $b \rightarrow q\tau\bar{\nu}_\tau$
inclusive weak decay including first order perturbative QCD corrections.
For $m_\tau=0$ the results of \cite{JK,CJ} are rederived

\noindent
(ii) after numerical integration the distribution $d\Gamma / dt$ is obtained 
in perfect agreement with the analytic result of \cite{CJK} which has been
derived in a completely different way

\noindent
(iii) when combined with the non-perturbative QCD corrections calculated 
in \cite{nonp1,nonp2,nonp3} the results of this article describe inclusive 
decays of beautiful hadrons into $\tau$~lepton and anything

\noindent
(iv) the energy spectrum and a few its moments are calculated for 
$B$~into~$\tau$ transitions.

\vspace{1em}
\par\noindent
{\large\bf Acknowledgments}\\
MJ would like to thank Wolfgang Hollik, Frans Klinkhamer and Hans K\"uhn
for their warm hospitality during his stay in Karlsruhe. A part of the
calculations presented in this article was done in Asper Center for Theoretical
Physics in summer 1995.
LM is very grateful to Professor Kacper Zalewski for numerous discussions
clarifying many points of the Heavy Quark Effective Theory.

%\newpage


\begin{thebibliography}{99}
\bibitem{VS}
M.~Voloshin and M.~Shifman, Sov.~J.~Nucl.~Phys.~45 (1987) 292; 47 (1988) 511.
\bibitem{PW}
 H.D.~Politzer and M.B.~Wise, Phys.~Lett.~B206 (1988) 681;
                                          B208 (1988) 504.
\bibitem{IsgW}
 N.~Isgur and M.B.~Wise, Phys.~Lett.~B232 (1989) 113; B237 (1990) 527.
\bibitem{EH}
E.~Eichten and B.~Hill, Phys.~Lett~B234 (1990) 511.
\bibitem{Grin}
B.~Grinstein, Nucl.~Phys.~B339 (1990) 253.
\bibitem{Geor}
H.~Georgi, Phys.~Lett.~B240 (1990) 447.
\bibitem{CGG}
J.~Chay, H.~Georgi and B.~Grinstein, Phys.~Lett. B247 (1990) 399.
\bibitem{IB}
I.I.~Bigi {\em et al.}, Phys.~Lett. B293 (1992) 430
[(E) {\em ibid.} B297 (1993) 477]; I.I.~Bigi {\em et al.},
Phys.~Rev.~Lett. 71 (1993) 496.
\bibitem{MWB}
A.V.~Manohar and M.B.~Wise, Phys.~Rev. D49 (1994) 1310;
              B.~Blok {\em et al.}, Phys.~Rev. D49 (1994) 3356
              [(E) {\em ibid.} D50 (1994) 3572;
              T.~Mannel, Nucl.~Phys. B413 (1994) 396.
\bibitem{MA}
G. Martinelli, plenary talk at ICHEP'96, July 1996, Warsaw, Poland;\\
G. Altarelli, summary talk at CRAD'96, August 1996, Cracow, Poland.
\bibitem{ALEPH2}
D. Buskulic et al. (ALEPH Collab.), Phys. Lett. B365 (1996) 437;
C. Diaconu,
{\it Measurement of the polarization  of the $\Lambda_b$ baryon in
$Z\to b\bar b$ event at the ALEPH experiment at LEP},
Provence U. Doctoral Thesis.
\bibitem{MeA}
B. Mele and G. Altarelli, Phys. Lett. B299 (1993) 345.
\bibitem{CJKK}
A. Czarnecki, M. Je\.zabek, J.G. K\"orner and J.H. K\"uhn,
                              Phys. Rev. Lett.~73 (1994) 317.
\bibitem{GR}
G. Bonvicini and L. Randall, Phys. Rev. Lett.~73 (1994) 392.
\bibitem{CKPS}
F.E. Close, J.G. K\"orner, R.J.N. Phillips and D.J. Summers,
                          J. Phys. G18 (1992) 1716.
\bibitem{PDG}
Particle Data Group Tables, R.M~Barnett et al., Phys. Rev.~D54 (1996)~1 and
 references therein.
\bibitem{CJK}
A.~Czarnecki, M.~Je\.{z}abek, H. K\"{u}hn, Phys. Lett. B~346 (1995) 335.
\bibitem{MV}
M.B.~Voloshin, Phys. Rev.~D51 (1995) 4934.
\bibitem{Kal}
J.~Kalinowski, Phys.Lett.~B245 (1990) 201.
\bibitem{GL}
Y.~Grossman and Z.~Ligeti, Phys.~Lett.~B332 (1994) 373.
\bibitem{GHN}
Y.~Grossman, H.E.~Haber and Y.~Nir,
              Phys.~Lett.~B357 (1995) 630.
\bibitem{nonp1}
A.F.~Falk, Z.~Ligeti, M.~Neubert and Y.~Nir, Phys. Lett~326 (1994) 145.
\bibitem{nonp2}
L.~Koyrakh, Phys. Rev.~D49 (1994) 3379; L.~Koyrakh, PhD. Thesis,
hep/ph9607443.
\bibitem{nonp3}
S.~Balk, J.G.~K\"{o}rner, D.~Pirjol and K.~Schilcher, Z.~Phys.~C64
(1994) 37.
\bibitem{HP}
Q. Hokim and X.-Y. Pham, Ann. Phys. 155 (1984) 202.
\bibitem{BBBG}
E. Bagan, P. Ball, V.M. Braun and P. Gosdzinsky, Nucl. Phys. B432 (1994) 3.
\bibitem{Boyd}
C.G.~Boyd, F.J.~Vegas, Z.~Guralnik and M.~Schmaltz UCSD/PTH 94--22, Dec.~1994
(unpublished).
\bibitem{JK}
M.~Je\.{z}abek, H.~K\"{u}hn, Nucl. Phys. B320 (1989) 20.
\bibitem{CJ}
A.~Czarnecki, M.~Je\.{z}abek, Nucl. Phys. B427 (1994) 3.
\bibitem{JM1}
M. Je\.zabek and L. Motyka, Acta Phys. Polon. B27 (1996) 3603.
\bibitem{Behrends}
R.~Behrends, R.~Finkelstein and A.~Sirlin Phys. Rev. 101 (1955) 866.
%S.~Berman, Phys.~Rev. 112 (1958) 267.
\bibitem{JK1}
M.~Je\.{z}abek, H.~K\"{u}hn, Nucl. Phys. B 314 (1989) 1.
\bibitem{CD}
A.~Czarnecki and S.~Davidson, Phys. Rev. D48 (1993) 4183.
\bibitem{GKLW}
M.~Gremm, A.~Kapustin,~Z.Ligeti and M.B.~Wise, Phys.Rev.Lett.~77 (1996) 20.
\bibitem{GK96}
M.~Gremm and A.~Kapustin, hep-ph/9603448 and CALT-68-2042 (1996). 
\end{thebibliography}
\end{document}